\documentclass[11pt]{article}
\usepackage{amsmath,amssymb,color,epsfig,cite}
\usepackage{graphicx}
\usepackage{subfigure}
\usepackage{setspace}

\usepackage{amsfonts}

\newcommand{\hoch}[1]{$\, ^{#1}$}

\makeatletter
\@addtoreset{equation}{section}
\makeatother

\newcommand{\be}{\begin{equation}}
\newcommand{\ee}{\end{equation}}
\newcommand{\bea}{\setlength\arraycolsep{2pt} \begin{eqnarray}}
\newcommand{\eea}{\end{eqnarray}}
\newcommand{\nn}{\nonumber}

\def\ft#1#2{{\textstyle{\frac{\scriptstyle #1}{\scriptstyle #2} } }}
\def\fft#1#2{{\frac{#1}{#2}}}

\def\0{{\sst{(0)}}}
\def\1{{\sst{(1)}}}
\def\2{{\sst{(2)}}}
\def\3{{\sst{(3)}}}
\def\4{{\sst{(4)}}}
\def\5{{\sst{(5)}}}
\def\6{{\sst{(6)}}}
\def\7{{\sst{(7)}}}
\def\8{{\sst{(8)}}}
\def\9{{\sst{(9)}}}

\def\sst#1{{\scriptscriptstyle #1}}

\begin{document}



\begin{center}
{\large {\bf Causality and $a$-theorem Constraints on Ricci Polynomial\\ and Riemann Cubic Gravities}}

\vspace{10pt}
Yue-Zhou Li$^1$\hoch{\dag},  H. L\"u$^1$\hoch{*} and Jun-Bao Wu$^{1, 2}$\hoch{\ddag}

\vspace{15pt}

{\it $^1$ Department of Physics, Tianjin University, Tianjin 300350, China\\
$^2$ Center for High Energy Physics, Peking University, Beijing 100871, China}

\vspace{30pt}

\underline{ABSTRACT}

\end{center}

In this paper, we study Einstein gravity extended with Ricci polynomials and derive the constraints on the coupling constants from the considerations of being ghost free, exhibiting an $a$-theorem and maintaining causality.  The salient feature is that Einstein metrics with appropriate effective cosmological constants continue to be solutions with the inclusion of such Ricci polynomials and the causality constraint is automatically satisfied.  The ghost free and $a$-theorem conditions can only be both met starting at the quartic order.  We also study these constraints on general Riemann cubic gravities.

\vfill {\footnotesize \hoch{\dag}liyuezhou@tju.edu.cn\ \ \ \hoch{*}mrhonglu@gmail.com\ \ \ \hoch{\ddag}junbao.wu@tju.edu.cn}

\pagebreak

\tableofcontents
\addtocontents{toc}{\protect\setcounter{tocdepth}{2}}


\newpage

\section{Introduction}

There has been a steady flow of works on Einstein gravity extended with higher-order curvature polynomial invariants, even more so in the light of the AdS/CFT correspondence \cite{Maldacena:1997re}.  With appropriate coupling constants, maximally-symmetric metrics with zero, positive or negative cosmological constants, corresponding to Minkowski, de Sitter (dS) and anti-de Sitter (AdS) spacetimes, can arise as the vacua. The renormalizability of such a theory associated with the higher power of the propagators in a vacuum turns out to conflict with the ghost-free condition \cite{Stelle:1976gc,Stelle:1977ry}.  In general, the linearized theory on the vacuum contains the massive scalar and ghost-like massive spin-2 modes, in addition to the usual graviton mode.  In four dimensions, for quadratically extended gravities, the absence of the ghosty massive spin-2 mode leads to the Starobinsky $R+R^2$ cosmological model \cite{Starobinsky:1980te} which contains a massive scalar mode. It was shown \cite{Nelson:2010ig,Lu:2015cqa} that there is no new static and spherically-symmetric black hole in the Starobinsky model besides the usual Schwarzschild metric.  New such a black hole does exist however in the theories with ghost massive spin-2 modes \cite{Lu:2015cqa,Lu:2015psa,Lu:2017kzi}.  In five dimensions and beyond, there exists a non-trivial Gauss-Bonnet combination
\be
L_{\rm GB} = R^2 - 4 R^{\mu\nu}R_{\mu\nu} + R^{\mu\nu\rho\sigma} R_{\mu\nu\rho\sigma}\,,
\ee
for which the linearized theory contains neither massive modes. The Schwarzschild black hole is replaced by the Schwarzschild-like one \cite{Boulware:1985wk,Cai:2001dz} where the static ansatz
\be
ds^2=-h(r) dt^2 + \fft{dr^2}{f(r)} + r^2 d\Omega^2\label{staticmet}
\ee
continues to be constrained by $h=f$, the same as the Schwarzschild metric.  The Ricci scalar and the Gauss-Bonnet combination belong to a series of Euler integrands of $k$'th orders which form the Lovelock gravities \cite{ll}.  The massive scalar and spin-2 modes are absent in these linearized theories.  However, the Lovelock term of $k$'th order becomes trivial at dimensions less and equal $D=2k$: it is topological and a total derivative at $D=2k$ and simply vanishes at $D<2k$.  Pure Lovelock gravities are thus irrelevant in lower dimensions.

The decoupling of both the massive scalar and ghosty spin-2 modes requires only two linear algebraic constraints on the coupling constants.  Ghost free theories with no massive scalar mode are thus not hard to come by beyond the quadratic order, even in four dimensions \cite{quasi0,quasi1,Sisman:2011gz,quasi2,quasi3,
Bueno:2016xff,Bueno:2016ypa,Cisterna:2017umf,Hennigar:2017ego,Ahmed:2017jod,quasi4}. Such theories exist even for cubic Ricci polynomial gravity in four dimensions \cite{quasi4}.  Ricci polynomial gravities up to and including the tenth order were studied in \cite{quasi4}.  There are two salient features.  The first is that the decoupling of both massive modes implies that the higher-order curvature terms give no contribution to the linearized gravity in general dimensions, as if these terms are topological.  The theories were called linearized quasi-topological gravity in \cite{quasi4}.  The second is that this property of the linearized theories extend to a generic Einstein metric background with appropriate effective cosmological constant rather than only on the maximally-symmetric vacuum spacetimes.

The massive scalar mode in the linear spectrum does not violate the ghost free condition.  Its exclusion comes also from an AdS/CFT consideration.  In even dimensional conformal field theories (CFT), conformal anomaly (also known as Weyl or trace anomaly) may arise such that the trace of the energy-momentum tensor becomes non-vanishing even for the vacuum,namely \cite{Duff:1977ay,Duff:1993wm,Imbimbo:1999bj}
\be
\langle T_\mu{}^\mu{}\rangle\sim -a E^{(2n)}+\sum_{i}c_{i}I^{(2n)}\,,
\ee
where $E^{(2n)}$ is the $n$'th order Euler integrand and $I_i$'s are the Weyl invariants in $D=2n$ dimensions. The constants $a$ and $c_i$ represent two different types of central charges. In planar ${\cal N}=1$, $D=4$ superconformal field theory, the two central charges are equal, namely $a= c$.  These central charges can be obtained from dual AdS$_5$ supergravities.  It is clear that for Ricci-polynomial gravities, where quadratic and/or higher-order Riemann tensor terms cannot be generated, and hence we have $c_i \propto a $.  This is supported by the fact that the ${\cal N}=1$, $D=4$ superconformal field theory is dual to some ${\cal N}=2$, $D=5$ gauged supergravity.
When gravities involve higher than linear Riemann tensor terms, the $a$ and $c_i$ central charges may become independent parameters.  These holographic anomalies are related to some holographic renormalization group (RG) procedure \cite{Girardello:1998pd,Freedman:1999gp,deBoer:2000cz,Skenderis:2002wp,Rajagopal:2015lpa}.  Many related works can be found in \cite{Henningson:1998gx,Nojiri:1999mh,Blau:1999vz,deHaro:2000vlm,Schwimmer:2003eq,
Banados:2004zt,Kraus:2005zm,Banados:2005rz,Banerjee:2009fm,Myers:2010jv,Sen:2012fc,Bugini:2016nvn}.
In this paper, we make use of the observation that metric on the round sphere is conformally flat.
This allows us to develop a simple technique to calculate the contributions to the $a$ charge from higher order curvature terms by using a reduced Fefferman-Graham (FG) expansion with the round spheres as the level surfaces.

    A remarkable result in $D=4$ CFT is the $a$-theorem, which was earlier conjectured in
\cite{Cardy:1988cwa}. A proof was recently proposed \cite{Komargodski:2011vj}. This generalizes the Zamolodchikov's $c$-theorem in $D=2$ \cite{zamo}.  The generalization to $D=6$ was also obtained in \cite{Cordova:2015vwa, Cordova:2015fha} for $(2, 0)$ and $(1, 0)$ theories, respectively. The $D=6$ $a$-theorem was also proved using another approach in \cite{Huang:2015sla}, based on \cite{Chen:2015hpa}.  The theorem states that in any RG flow connecting two conformal fixed points, one ultra-violate (UV) and one infra-red (IR), with
\be
a_{\rm UV} \ge a_{\rm IR}\,,
\ee
the central charge $a$ is monotonously increasing from the IR region to the UV region.  In the context of the AdS/CFT correspondence where energy is dual to the bulk radius $r$, this statement becomes
\be
a'(r) \ge 0\,,\qquad a(r)\Big|_{\rm AdS}=a\,,
\ee
for certain $a(r)$ function.  (Here we assume that $r\rightarrow \infty$ is the asymptotic AdS boundary, corresponding to the UV region of the dual CFT.) It turns out that the holographic $a$-theorem in $D=5$, ${\cal N}=8$ gauged supergravity can be related to the null-energy condition (NEC) in domain metrics \cite{Freedman:1999gp}.  This provides a mean of generalizing the holographic $a$-theorem to general odd  dimensions \cite{Myers:2010xs}.  This approach is non-covariant and covariant approaches were also proposed in, e.g.~\cite{Sahakian:1999bd,Anber:2008js}.   It is clear that the NEC can be also studied in even dimensions and analogous results can be obtained.  However, in even dimensions there is no conformal anomaly. The corresponding charge $a=a(r)|_{\rm AdS}$ was instead interpreted as a universal finite piece in  entanglement entropy for a spherical entangling surface \cite{Myers:2010tj} and sphere partition functions \cite{Casini:2011kv}. (It should be clarified that owing to historical reasons, the $a$-function was many times in literature referred as the $c$-function and the related theorem in three dimensions was also referred as $F$-theorem \cite{Jafferis:2011zi, Klebanov:2011gs} due to its relation to sphere partition functions.)

In this paper, we follow \cite{Myers:2010xs,Myers:2010tj} and determine a natural class of holographic $a$-functions, based on the NEC on domain walls, assuming that matter is minimally coupled.  We devise a simple procedure to relate the $a$-theorem to the NEC, by introducing a free scalar field.  We find that the validation of the $a$-theorem requires the decoupling of the massive scalar modes; however, the massive spin-2 mode does not affect the $a$-theorem even though it is ghost-like.  Thus our $a$-theorem constraint together with the ghost-free condition imply that both massive modes should be absent in Einstein gravity extended with higher-order curvature polynomial invariants.

Recently by studying the scattering process of gravitons, further causality constraints were discovered \cite{Camanho:2014apa}.  To be specific, for two-derivative linearized gravities, the higher-order contribution to the linearized equation can be parameterized as
\be
\delta R_{\mu\nu} + \alpha_2 \check R_{(\mu}{}^{\rho\alpha\beta} \delta R_{\nu)\rho\alpha\beta} +
\ft12\alpha_4 [\nabla_{(\mu} \nabla_{\nu)} \check R^{\alpha\beta\rho\sigma}] \delta R_{\alpha\beta\rho\sigma}=0\,,
\ee
where the checked quantity is the subtracted Riemann tensor \cite{Camanho:2014apa}.  It was demonstrated that for non-vanishing $\alpha_2$ and $\alpha_4$, causality will be violated \cite{Camanho:2014apa}.  Thus $\alpha_2$ and $\alpha_4$ should be zero when gravity is treated as a complete theory. Consequently, the high-order terms must be at least quasi-topological at the linear level. On the other hand, when gravity is viewed as an effective theory embedded in a bigger and more fundamental theory e.g.~strings, for sufficiently small $\alpha_2$ and $\alpha_4$, new degrees of freedom such as higher-spin modes may come to rescue in the region where causality could be potentially violated \cite{Camanho:2014apa}.  In either cases, causality provides very strong constraints on the coupling constants of the higher-order terms such as Lovelock gravities.  In the Ricci polynomial gravities considered in \cite{quasi4} and in this paper, for perturbations around the Einstein metric solutions, the $\alpha_2$ and $\alpha_4$ term will not arise.  As we have mentioned earlier, after decoupling the massive scalar and spin-2 modes, the theories become quasi-topological in that the linearized gravity is identical to that of Einstein gravity.  The causality constraint is thus trivially satisfied.  From the AdS/CFT perspective, this constraint reflects that the $a$ and $c$ charges are equal in the four-dimensional CFT, as in the case of the planar $D=4$, ${\cal N}=1$ superconformal field theory  \cite{Camanho:2014apa}.  Indeed for Ricci polynomial gravities, we do have $a=c$.  We demonstrate this also for general Riemann cubic terms, and show that $a=c$ is equivalent to the condition for linearized quasi-topological gravities, after massive modes are decoupled.

The paper is organized as follows.  In section 2, we study Einstein gravity with a bare cosmological constant, extended with three Ricci cubic terms.  We compute the holographic central charge in five dimensions and demonstrate $a=c$ indeed.  We then construct a natural $a$-function and derive the corresponding $a$-theorem.  We show that the validity of the $a$-theorem requires two linear constraints on the coupling constants, one of which implies the decoupling of the massive scalar mode.  We then show that the $a$-theorem constraints are inconsistent with the ghost free condition for Ricci cubic gravity.  We then generalize the results to general dimensions.  In section 3, we consider higher-order Ricci polynomials and construct ghost free theories that are also consistent with the $a$-theorem constraints.  We find that such theories arise at the quartic order. Starting at the sixth order, ghost free theories satisfying the $a$-theorem constraints can also be quasi-topological on both special and general static metrics, as defined in \cite{quasi4}.   In section 4, we consider the general Riemann cubic theories and study the ghost free, $a$-theorem and casuality constraints in five and general dimensions.  The paper is concluded in section 5.  In appendix A, we present the Riemann curvature tensor for a class of domain wall metrics that will be extensively used in this paper.

\section{Ricci cubic gravities}

\subsection{The theory and the linear spectrum}

In this section, we consider Einstein gravity coupled to a bare negative cosmological constant $\Lambda_0$, extended with three Ricci cubic invariants.  The Lagrangian is given by
\be
{\cal L} = \sqrt{-g}\, (R - 2\Lambda_0 + L^\3)\,,\qquad
L^\3 = e_1 R^3 + e_2 R\,R_{\mu\nu} R^{\mu\nu} + e_3 R^{\mu}_{\nu} R^{\nu}_\rho R^{\rho}_\mu\,.\label{cubiclag}
\ee
The theory was studied in detail in \cite{quasi4}. The covariant equation of motion associated with the variation of the metric $\delta g^{\mu\nu}$ is
\be
{\cal E}_{\mu\nu} \equiv  P_{\mu \alpha\beta\gamma}R_\nu{}^{\alpha\beta\gamma}-\ft12g_{\mu\nu}(R-2\Lambda_0+ L^\3)-2\nabla^\alpha\nabla^\beta P_{\mu\alpha\beta\nu} = 0\,,\quad (P_{\mu\nu\rho\sigma} \equiv \fft{\partial L}{\partial R^{\mu\nu\rho\sigma}}\,.)\label{geneom}
\ee
Einstein metrics with appropriate effective cosmological constants are solutions of the theory, namely
\be
\bar R_{\mu\nu} = \fft{2\Lambda_{\rm eff}}{D-2} \bar g_{\mu\nu}\,,
\ee
where $\Lambda_{\rm eff}$ satisfies the cubic algebraic equation
\be
\Lambda_0 =\Lambda_{\rm eff} + \ft{4(D-6)}{(D-2)^3} (D^2 e_1 + D e_2 + e_3) \Lambda_{\rm eff}^3\,.
\label{Deffeom}
\ee
The linearized spectrum around the Einstein metrics contains the usual graviton mode.  In addition, there are massive scalar and spin-2 modes, with the masses \cite{quasi4}
\bea
\mu_0^2 &=& -\frac{3 (D-6) (D-1)^2 \left(D \left(D e_1+e_2\right)+e_3\right)+(D+2) \ell^4}{(D-1) \left(D \left(12 (D-1) e_1+(D+8) e_2+3 e_3\right)-8 e_2\right)\ell^2}\,,\nn\\
\mu_2^2 &=& \frac{(D-1)^2 \left(9 D^2 e_2+13 D e_2+21 e_3\right)+\ell^4}{(D-1)\left(D e_2+3 e_3\right)\ell^2 }\,.\label{mu0mu2}
\eea
Here $\ell$ is the radius of AdS, defined by
\be
\Lambda_{\rm eff} = -\fft{(D-1)(D-2)}{2\ell^2}\,.\label{lameff}
\ee
It is worth pointing out here that general Einstein metrics continue to be solutions even if we include a linear Riemann tensor term such as $R^{\mu\nu} R_{\rho\sigma} R_{\mu\rho\nu\sigma}$.  We shall discuss this term in section \ref{sec:riemann}.

It was well known that the massive spin-2 mode is ghost like with negative kinetic term.  The ghost-free condition at the linear level requires the decoupling of the ghost mode from the spectrum, i.e.
\be
De_2 + 3e_3=0\,.\label{cubicghostfree}
\ee
The absence of both massive modes yields
\be
{\cal W}^\3=R^3 - \ft32D R\,R_{\mu\nu} R^{\mu\nu} + \ft12 D^2 R^{\mu}_{\nu} R^{\nu}_\rho R^{\rho}_\mu\,.
\ee
It turns out that this cubic combination becomes quasi-topological at the linear level in that it gives no contribution to the linearized equations of motion \cite{quasi4}, and thus satisfying the casuality condition of \cite{Camanho:2014apa}. Furthermore, it was observed \cite{quasi4} that in $D=4$, it becomes quasi-topological on special static metrics (\ref{staticmet}) with $h=f$.

Note that the ghost free condition only requires the decoupling of the massive spin-2 mode.  The absence of the massive scalar mode is required also by an $a$-theorem, which we shall discuss next.  Furthermore, as mentioned earlier, the absence of both massive modes are required by the causality condition so that the resulting theory becomes linearly quasi-topological.

\subsection{Central charge and an $a$-theorem in $D=5$}

\subsubsection{Holographic anomaly and central charge}

In this subsection, we focus our discussion on five dimensions.  We first consider
the FG expansion in general $D=d+1$ dimensions:
\begin{equation}
ds^{2}=\dfrac{\ell^2}{4\rho^{2}}d\rho^{2}+\dfrac{1}{\rho}g_{ij}dx^{i}dx^{j}\,,
\end{equation}
where
\begin{equation}
\begin{split}
&g_{ij}=g^{\0}_{ij}+\rho g^{\2}_{ij}+\rho^{2}g^{\4}_{ij} + \cdots
\\&g^{ij}=g^{\0 ij}-\rho g^{\2 ij}-\rho^{2}(g^{\4 ij}-\ft{1}{4}{\rm Tr}(g^{\2 2})g^{\0 ij})+ \cdots\,.
\end{split}
\end{equation}
For our Ricci polynomial gravities, we only need consider the Ricci tensors. The relevant components are
\begin{equation}
\begin{split}
&R_{ij}=\hat{R}_{ij}-\dfrac{2\rho}{\ell^2}g''_{ij}+\dfrac{1}{\ell^2}g^{lk}g'_{lk}g_{ij}+
\dfrac{2\rho}{\ell^2}g^{kl}g'_{lj}g'_{ki}
-\dfrac{\rho}{\ell^2}g^{lk}g'_{lk}g'_{ij}+\dfrac{d-2}{\ell^2}g'_{ij}-\dfrac{d}{\ell^2}
\dfrac{1}{\rho}g_{ij}\,,
\\&R_{\rho\rho}=-\dfrac{d}{4\rho^{2}}-\dfrac{1}{2}g^{ij}g''_{ij}+
\dfrac{1}{4}g^{ik}g^{jl}g'_{ij}g'_{kl}\,,
\end{split}
\end{equation}
where a prime denotes a derivative with respect to $\rho$.  Substituting all these into the action for the Lagrangian (\ref{cubiclag}), we find in five dimensions
\begin{equation}
S=\dfrac{1}{16\pi}\int d^{4}x\int_{\epsilon} d\rho \sqrt{-g} {\cal L} = \fft{1}{16\pi} \int
d^4 x \int d\rho \Big(\cdots + \fft{a_\4}{\rho} + \cdots\Big)\,,
\end{equation}
where the coefficient $a_\4$ is given by
\begin{eqnarray}
a_{\4} &=& \alpha_{1}\hat{R}^{(0)2}+\alpha_{2}\hat{R}^{(0)}_{ij}\hat{R}^{(0)ij}+\beta {\rm Tr} g^{(4)}+A \big({\rm Tr} (g^{(2)2})+\ell^2 \hat{R}^{(0)}_{ij}g^{(2)ij}\big)\nn\\
&&+B({\rm Tr} g^{(2)})^{2}
+C\hat{R}^{(0)}{\rm Tr} g^{(2)}\,.
\end{eqnarray}
Various coefficients above are given by
\begin{eqnarray}
&&\alpha_{1}=-\dfrac{1}{\ell^2}(30e_{1}+4e_{2})\,,\qquad \alpha_{2}=\dfrac{1}{\ell^2}(10e_{2}+6e_{3})\,,\cr
&&\beta=-\dfrac{1}{2\ell^{6}}(6\ell^{4}-32(25e_{1}+5e_{2}+e_{3})+ \ell^{6}\Lambda_0)\,,\quad
A=-(\dfrac{1}{2\ell^{2}}+\dfrac{600e_{1}}{\ell^{6}}+\dfrac{160e_{2}}{\ell^{6}}+
\dfrac{48e_{3}}{\ell^{6}})\,,\cr
&&B=\dfrac{1}{2\ell^{2}}-\dfrac{480e_{1}}{\ell^{6}}-\dfrac{104e_{2}}{\ell^{6}}
-\dfrac{24e_{3}}{\ell^{6}}\,,\quad C=\dfrac{1}{4}-\dfrac{60e_{1}}{\ell^{4}}-\dfrac{8e_{2}}{\ell^{4}}\,.
\end{eqnarray}
It follows from (\ref{Deffeom}) that $\beta=0$. The remaining terms then depend only on $g^{(2)}_{ij}$. Equations of motion associated with the variation of $g^{(2)}_{ij}$ yield
\begin{equation}
{\rm Tr} g^{(2)}=-\dfrac{4C+A\ell^2}{2(A+4B)}\hat{R}^{(0)}\,,\qquad
g^{(2)}_{ij}=\dfrac{B\ell^2-C}{2(A+4B)}g^{(0)}_{ij}-\ft12\ell^2\hat{R}^{(0)}_{ij}\,.
\end{equation}
Substituting these back into the action, we have
\begin{equation}
a_{\4}=\Big(\dfrac{\ell^{3}}{8}+\dfrac{6}{\ell}(25e_{1}+5e_{2}+e_{3})\Big)
(\hat{R}^{(0)}_{ij}\hat{R}^{(0)ij}-\ft{1}{3}\hat{R}^{(0)2})\,.\label{d5riccianomaly}
\end{equation}

In general, the gravitational anomaly in four dimensions contains both $a$ and $c$ charges, given by
\be
a_\4= \sqrt{-h} (c I^\4 - a E^\4)\,,
\ee
where $I^\4$ is Weyl tensor squared and $E^\4$ is the Euler integrand (Gauss-Bonnet term) in four dimensions, namely
\bea
I^\4 &=& C^{\mu\nu\rho\sigma} C_{\mu\nu\rho\sigma} = R^{\mu\nu\rho\sigma} R_{\mu\nu\rho\sigma} - 2 R^{\mu\nu} R_{\mu\nu} + \ft13 R^2\,,\nn\\
E^\4 &=& \delta^{a_1b_1a_2b_2}_{c_1d_1c_2d_2} R^{c_1d_1}{}_{a_1b_1} R^{c_2d_2}{}_{a_2d_2} = R^{\mu\nu\rho\sigma} R_{\mu\nu\rho\sigma} - 4 R^{\mu\nu} R_{\mu\nu} + R^2\,.
\eea
The absence of Riemann tensor in the anomaly (\ref{d5riccianomaly}) implies that $c=a$, and furthermore we have
\begin{equation}
c=a=\dfrac{\ell^{3}}{8}+\dfrac{6}{\ell}(25e_{1}+5e_{2}+e_{3})\,.
\label{cubica}
\end{equation}
In this paper, we shall not be fastidious on the overall pure numerical $\pi$ factors of the central charges.  As we shall see presently, the $a=c$ result is closely related to the fact the Ricci polynomials are linearly quasi-topological after massive modes are decoupled.

\subsubsection{An $a$-function and the corresponding $a$-theorem}

We follow \cite{Myers:2010xs} and consider the cohomogeneity-one domain wall metric ansatz
\be
ds^2 = dr^2 + e^{2A(r)} (-dt^2 + dx_1^2 + dx_2^2 + dx_3^2)\,,\label{ads5dw}
\ee
The AdS spacetime is given by $A(r)=r/\ell$, where $\ell$ is the AdS radius, with the AdS boundary located at $r\rightarrow \infty$.  The function $e^{2A}$ thus describes the flow to the AdS with $\ell=1/A'$ at the AdS ``fixed'' point.  We can substitute $\ell=1/A'$ into the $a$-charge (\ref{cubica}) and define an $a$-function
\be
a(r)=\dfrac{1}{8A'^3}+6(25e_ 1+5e_ 2+e_ 3) A'\,,
\ee
which implies that
\be
a'(r)=\ft{3}{8}\Big(-\dfrac{1}{A'^4}+16(25e_ 1+5e_ 2+e_ 3)\Big)A''\,.\label{cubicaprime}
\ee
The $a$-theorem implies that $a'\ge 0$.  In Einstein gravity with $e_i=0$, it can be easily verified that
\be
-T^{t}_{t} + T^{r}_{r}=3 A'' = 8 A'^4\, a'\,.
\ee
The NEC, i.e. $(-T^{t}_t + T^{r}_{r}) \ge 0$ thus implies indeed
\be
a'\ge 0\,.
\ee

The situation with non-vanishing cubic terms is more complicated.  From the equations of motion, we find
\bea
&&-T^{t}_{t} + T^{r}_r = -{\cal E}^{t}_{t} + {\cal E}^{r}_r=8 A'^4\, a' -3 \Big[\left(128 e_1+40 e_2+17 e_3\right)(A'''^2+
A''A'''')\nn\\
&&+4 \left(208 e_1+59 e_2+22 e_3\right) A''^3+4\left(448 e_1+116 e_2+37 e_3\right) A' A'' A'''\nn\\
&&+4  \left(80 e_1+19 e_2+5 e_3\right) (4A'^3A'''+ A'^2 A''''+16 A'^2 A''^2)\Big]\,,\label{cubicd5nec}
\eea
where $a'$ is given by (\ref{cubicaprime}). For general $e_i$ coupling constants, the function $a(r)$ is not monotonic function of $r$.  Thus at least for the choice of our $a$-function, albeit natural, the $a$-theorem breaks down for general $e_i$ couplings. However, if we require that the couplings satisfy
\be
80 e_1+19 e_2+5 e_3=0\,,\qquad 128 e_1+40 e_2+17 e_3=0\,,\label{cubicacon}
\ee
we have
\be
\fft8{\pi} A'^4\, a' =-T^{t}_{t} + T^{r}_r \ge 0\,,
\ee
under the NEC for the matter energy-momentum tensor, and consequently the $a$-theorem is restored.

Note that the first term in (\ref{cubicacon}) implies that $\mu_0$ in (\ref{mu0mu2}) diverges and hence the massive scalar mode decouples from the theory.  However, the two constraints in (\ref{cubicacon}) for the $a$-theorem does not require the decoupling of the ghost-like massive spin-2 mode.  In fact, they are inconsistent with ghost-free condition (\ref{cubicghostfree}). For Ricci polynomials, we have to consider higher orders to satisfy both the $a$-theorem and ghost free conditions.  We shall discuss this in section 3.   If we also include Riemann polynomials, the ghost free and $a$-theorem conditions can be both met at the cubic order, but the causality condition cannot be met within the gravity theory.  We shall address this in section 4.  In the next subsection, we generalize our results to general dimensions.

\subsection{Generalizing to higher dimensions}

\subsubsection{A simple method for deriving the NEC}
\label{sec:simplenec}

In order to establish the conditions on the coupling constants of the Ricci polynomials for the $a$-theorem, it is necessary to perform two computations.  The first is to compute the contributions to the central charges from the higher-order terms.  This enables us to determine the $a$ charge and hence define a corresponding $a$-function. The second is to relate the derivative of the $a$-function with the NEC.

In this subsection, we develop a simple strategy of deriving the NEC, which implies that
\be
-{\cal E}^{t}_t + {\cal E}^r_r\ge 0\,,
\ee
for the domain wall metrics. Even though the curvature components for the domain wall metric ansatz can be easily obtained, the evaluation of ${\cal E}_{\mu\nu}$ in general dimensions can be involved.  Instead of considering the covariant equations of motion (\ref{geneom}) for domain walls, it is consistent and more convenient to substitute the ansatz into the Lagrangian and then derive the equations by the variational principle on the reduced fields.  However, this makes it harder to read off the combination $-{\cal E}^{t}_t + {\cal E}^r_r$ from the equations. We employ the trick of coupling Einstein-Ricci cubic gravity to a free scalar field $\phi$, namely
\be
{\cal L} = \sqrt{-g} \big( L^{\rm gr} - \ft12 (\partial\phi)^2\big)\,.\label{scalarlag}
\ee
For the domain wall metric ansatz, we assume that $\phi=\phi(r)$. It can be easily established that scalar energy-momentum tensor satisfies
\be
-T^{t}_t + T^{r}_{r} = \ft12 \phi'^2\,.
\ee
Einstein equations of motion then implies
\be
\ft12\phi'^2 = -{\cal E}^{t}_t + {\cal E}^r_r\,.
\ee

To be concrete, we consider the domain wall ansatz in general $D$ dimensions
\be
ds_D^2 = e^{2B(r)} dr^2 + e^{2A(r)} dx^\mu dx^\nu \eta_{\mu\nu}\,,\qquad
\phi=\phi(r)\,.\label{genddw}
\ee
The non-vanishing Ricci tensor components are
\be
R^r_r = (D-1) e^{-2B} \Big(-A'' + A' B' -A'^2\Big)\,,\qquad
R^\mu_\nu = e^{-2B} \Big(-A'' + A'B' -(D-1) A'^2\Big)\delta^\mu_\nu\,.
\ee
Substituting the ansatz into the Lagrangian (\ref{scalarlag}) where $L^{\rm gr}$ is given by (\ref{cubiclag}), we obtain the reduced effective Lagrangian of fields $(A,B,\phi)$.  Varying $(A,B)$ yields two nonlinear differential equations of $(A,B)$.  Eliminating the bare cosmological constant $\Lambda_0$, we obtain an equation involving a term $\phi'^2$, which we can now solve straightforwardly.  Making a coordinate gauge $B=0$, we find
\bea
\ft12\phi'^2 &=& -{\cal E}^{t}_t + {\cal E}^r_r\nn\\
&=&\Big(-(D-2) -3 (D-6) (D-1)^2 (D^2 e_1 + D e_2 + e_3)A'^4\Big)A''\nn\\
&&-(D-1) x \left(A''''+(D-1) (4A''^2 + A' A''')\right) A'^2-3 y (A'''^2+A''A'''')\nn\\
&&-(D-1) \left((x+3 y) A''^3+ (4 x+3 y) A' A''A'''\right)\,,\label{cubicgendnec}
\eea
where the coefficients $(x,y)$ are
\bea
x &=& 12 (D-1) D e_1+\left(D^2+8 D-8\right) e_2+3 D e_3\,,\nn\\
y &=& 8 (D-1)^2 e_1+2 D (D-1) e_2+\left(D^2-2 D+2\right) e_3\,.
\eea
Note that when $D=5$, (\ref{cubicgendnec}) becomes precisely (\ref{cubicd5nec}).  We now impose the condition that the expression $-{\cal E}^{t}_t + {\cal E}^r_r$ does not involve higher derivative terms $A'''$ and $A''''$.  This can be achieved if we choose the coupling constants $e_i$ such that
\be
x=0\,,\qquad y=0\,.\label{cubicgendquasicond}
\ee
Note that the $x$ combination of the coupling constants is the denominator of $\mu_0^2$ in (\ref{mu0mu2}). Thus $x=0$ implies that decoupling of the massive scalar mode.  The resulting cubic polynomial in $D$ dimensions, after imposing (\ref{cubicgendquasicond}), is
\be
H^\3 = (D^2+4 D-4)R^3-12 D(D-1)R R^{\mu\nu}R_{\mu\nu} +16 (D-1)^2 R^{\mu}_{\nu} R^{\nu}_\rho R^{\rho}_\mu\,.
\ee
The NEC for the cubic-extended theory with $\lambda H^\3$ on the domain wall metrics now becomes
\be
-(D-2) \Big(1 + 3 \lambda (D-6) (D-2)^3 (D-1)^2 A'^4\Big) A''\ge 0\,.
\ee

\subsubsection{Central charge and the $a$ theorem}

The central charge calculation in higher dimensions can be complicated; however, for Ricci polynomial gravities, the situation becomes much simpler. In general $D=2n$ dimensions, there are multiple conformal invariant terms $I^{(2n)}_i$ \cite{Bonora:1985cq}, and the trace anomaly of the CFT in $D=2n$ dimensions
takes the form
\be
\langle T \rangle \sim -a E^{(2n)} + \sum_i c_i I_i^{(2n)}\,,
\ee
where the $a$ charge is associated with the Euler integrand $E^{(2n)}$.  The absence of the quadratic or higher order Riemann tensor terms implies $c_i$ can be all expressed in terms of $a$.  For example in four dimensions, we have seen that $c=a$.  In six dimensions, there are three conformal invariants
\bea
&& I_{1}=C_{\mu\nu\rho\sigma} C^{\mu\alpha\beta\sigma}C_\alpha{}^{\nu\rho}{}_{\beta}\,,\qquad
I_{2}=C_{\mu\nu\rho\sigma}C^{\rho\sigma\alpha\beta}C_{\alpha\beta}{}^{\mu\nu}\,,
\cr
&& I_{3}=C_{\mu\rho\sigma\lambda}(\delta^{\mu}{}_\nu{}\Box+4R^{\mu}_\nu{}-\ft{6}{5}
R\delta^{\mu}_\nu{})C^{\nu\rho\sigma\lambda}+\nabla_{\mu}J^{\mu}\,.
\eea
The identity
\be
E^\6 = 48 Q^\6 + 96 I^\6_1 + 24 I^\6_2 - 8 I^\6_3
\ee
implies that \cite{Bugini:2016nvn}
\be
c_1=96a\,,\qquad c_2=24 a\,,\qquad c_3=-8a\,.
\ee
Most general static and spherically-symmetric black holes in six-dimensional conformal gravity ${\cal L}=\sqrt{-g} Q^\6$ was constructed in \cite{Lu:2013hx}.

For our purpose of studying the $a$-theorem, it is not necessary to consider the general FG expansion. Instead, we can adopt the Euclidean signature and consider the spherical domain-wall ansatz
\be
ds^2_D = e^{2B(r)} dr^2 + e^{2A(r)} d\Omega_{2n}^2\,.\label{spheredw}
\ee
For this ansatz, $I^{(2n)}_i$ on the boundary all vanishes since $d\Omega_{2n}^2$ is conformally flat, whilst $E^{(2n)}$ is non-vanishing. It follows that only the $a$ charge survives in the action. Specifically, the non-vanishing Ricci tensor components are
\bea
R^r_r &=& (D-1) e^{-2B} \Big(-A'' + A' B' -A'^2\Big)\,,\nn\\
R^i_j &=& \Big(e^{-2B} (-A'' + A'B' -(D-1) A'^2) + (D-2) e^{-2B}\Big)\delta^{i}_j\,.
\eea
The reduced FG expansion corresponds to
\be
e^{2B}=\fft{\ell^2}{4r^2}\,,\qquad e^{2A}=\fft{f(r)}{r}\,,
\ee
where the AdS boundary is located at $r=0$ and $f$ is given by
\be
f=f_0 + f_2 r + f_4 r^2 + f_6 r^3 + \cdots\,.
\ee
Here $f_i$ are constant variables.  Substituting the ansatz into the Lagrangian (\ref{cubiclag}) and perform the small-$r$ expansion.  We find that the anomalous $1/r$ coefficient vanishes in even $D$ dimensions.  In $D=2k+1$ dimensions, the anomaly term is a function of $(f_2,f_4,\ldots, f_{2k})$.  In particular, $f_{2k}$ is linear and its variation gives rise to (\ref{Deffeom}) with $\Lambda_{\rm eff}$ given in (\ref{lameff}).  The variation of $(f_2,f_4,\ldots, f_{2k-2})$ implies that
\be
f_2=-\ft12\ell^2\,,\qquad f_4=\fft{\ell^4}{16 f_0}\,,\qquad f_{2i}=0\,,\qquad 3 \le i\le k-1\,.
\ee
We can thus read off the anomalous coefficient, which is proportional to the $a$ charge
\be
a=\ell^{D-2} + 3(D-1)^2 (D^2 e_1 + D e_2 + e_3) \ell^{D-6}\,.
\ee
This naturally suggests an $a$-function
\be
a=\fft{1}{A'^{D-2}} + 3(D-1)^2 (D^2 e_1 + D e_2 + e_3) A'^{6-D}\,.
\ee
Thus we have
\be
a'= \Big(-(D-2) - 3 (D-6)(D-1)^2 (D^2 e_1 + D e_2 + e3) A'^4\Big) \fft{A''}{A'^{D-1}}\,.
\ee
We see that $a'$ is exactly proportional to the first term in (\ref{cubicgendnec}) and $A'^{D-1}$ is non-negative in for odd $D$.  Thus the condition (\ref{cubicgendquasicond}), together with the NEC,  ensures the $a$-theorem.  However, the $a$-theorem condition and ghost free condition cannot be both satisfied at the cubic order for Ricci polynomials and we shall consider higher order terms in the next section.

\section{Higher-order Ricci polynomials}
\label{sec:higher-order}

General Ricci polynomial invariants are constructed from Ricci scalar $R$ and irreducible Ricci tensor polynomials.  The irreducible Ricci polynomial of $k$'th order is
\be
R_{(k)}=R^{\mu_1}_{\mu_2} R^{\mu_2}_{\mu_3} \cdots R^{\mu_k}_{\mu_1}\,.
\ee
The Ricci scalar can be viewed as $R=R_\1$.  The general Ricci polynomials at the $k$'th order can be expressed as
\be
{\cal R}_{icci}^{(k)} = e_1 R_\1^k + e_2 R^{k-2} R_\2 + e_3 R_\1^{k-3} {R}_\3 +
e_4 R_\1^{k-4} {R}_\2^2 +  e_5 R_\1^{k-4} {R}_\4 + \cdots\,.
\ee
Here we use a lexical ordering of the coupling constants, as in \cite{quasi4}.  The term with higher power of ${R}_{(k)}$ and with smaller $k$ has a smaller labelling index for its coupling constant.  In this paper, we shall work up to the tenth order of Ricci polynomials. We follow the same technique of section 2 and
obtain the independent number of Ricci polynomials ${\cal A}_k$ that are consistent with the relevant $a$-theorem. The numbers of possible terms for a given order $k$ are summarized in Table 1.
\bigskip
\begin{table}[ht]
\begin{center}
\begin{tabular}{|c|c|c|c|c|c|c|c|c|c|c|}
  \hline
  $k$ & 1 & 2 & 3 &4 &5 &6 &7 &8 &9 &10  \\ \hline
  $N_k$ & 1 & 2 & 3 &5 & 7 & 11 & 15 & 22 & 30 &42\\ \hline
   ${\cal A}_k$ & 1 & 1 & 1 &2 & 3 & 6 & 9 & 15 & 22 &33\\ \hline
${\cal A}_k^{\rm gf}$ & 1 & 0 & 0 &1 & 2 & 5 & 8 & 14 & 21 &32\\
  \hline
\end{tabular}
\caption{\small \it $N_{k}$ is the number of all possible Ricci polynomial terms at the $k$'th order. ${\cal A}_k$ denotes the number of independent combinations that are consistent with the $a$-theorem. ${\cal A}_k^{\rm gf}$ are those which also satisfy the ghost free condition.}
\end{center}
\end{table}
Here we give a few explicit low-lying examples
\bea
k=1:&& {\cal A}_\1 = R_\1\,,\cr
k=2:&& {\cal A}_\2 = D R_\1^2 - 4(D-1) R_\2\,,\cr
k=3:&& {\cal A}_\3 =(D^2+4 D-4)R_\1^3-12 D(D-1)R_\1 R_\2 +16 (D-1)^2 R_\3\,,\cr
k=4:&& {\cal A}_\4 =e_1 \Big[\left(D^2+4 D-4\right) R_\1^4-12 (D-1) \left(\left(D^2+D-1\right) R_\2^2-(D-1) D R_\4\right)\cr
&&\quad\qquad\qquad+16 (D-1)^2 R_\3 R_\1\Big]\cr
&&\qquad\quad+e_2 \Big[D^3 \left(R_\4-R_\2^2\right)+D^2 \left(15 R_\4-13 R_\2^2\right)+2 \left(D^2+4 D-4\right) R_\1^2 R_\2\cr
&&\qquad\qquad\quad-4 (D-1) D R_\1 R_\3+D \left(13 R_\2^2-32 R_\4\right)+16 R_\4\Big]\,,\cr
k=5:&& {\cal A}_\5 = -e_1 \Big[-16 (D-1)^2 \big(5 D^3 \left(R_\2 R_\3-R_\5\right)+D^2 \left(13 R_\5-10 R_\2 R_\3\right)\cr
&&\qquad\qquad\quad+2 D \left(5 R_\2 R_\3-8 R_\5\right)+8 R_\5\big)-\left(D^4+6 D^3-14 D^2+16 D-8\right) R_\1^5\cr
&&\qquad\qquad\quad+20 (D-1) R_\1 \left(\left(D^4+D^3-5 D^2+8 D-4\right) R_\2^2-(D-1) D^3 R_\4\right)\Big]\cr
&&\qquad\quad+e_2 \Big[2 \left(D^4+6 D^3-14 D^2+16 D-8\right) R_\2 R_\1^3 + D^5 \left(R_\4-R_\2^2\right) R_\1\cr
&&\qquad\qquad\quad- 4 (D-1) \left(D^4-14 D^3+6 D^2+16 D-8\right) R_\2 R_\3\cr
&&\qquad\qquad\quad+ 4 (D-1)D \left(D^3-13 D^2+24 D-12\right) R_\5-3 D^4 \left(7 R_\2^2-5 R_\4\right) R_\1\cr
&&\qquad\qquad\quad+D^3 \left(9 R_\2^2-80 R_\4\right) R_\1+8 D^2 \left(3 R_\2^2+20 R_\4\right) R_\1\cr
&&\qquad\qquad\quad-12 D \left(R_\2^2+12 R_\4\right) R_\1+48 R_\1 R_\4\Big]\cr
&&\qquad\quad+e_3 \Big[2 \left(D^4+6 D^3-14 D^2+16 D-8\right) R_\3 R_\1^2-2 D^5 \left(R_\2 R_\3-R_\5\right)\cr
&&\qquad\qquad\quad+ \left(D^4-9 D^3-19 D^2+56 D-28\right) R_\1 R_\2^2-2 D^4 \left(6 R_\2 R_\3-7 R_\5\right)\cr
&&\qquad\qquad\quad-5 D \left(D^3+3 D^2-8 D+4\right) R_\1 R_\4+12 D^3 \left(9 R_\2 R_\3-8 R_\5\right) \cr
&&\qquad\qquad\quad-16 D^2 \left(12 R_\2 R_\3-13 R_\5\right)+96 D \left(R_\2 R_\3-2 R_\5\right)+64 R_\5\Big]\,.
\eea

Imposing the ghost-free condition reduces the number of allowed combinations by 1.  The numbers are given in Table 1.  The low-lying examples are
\bea
{\cal A}^{\rm gf}_\4 &=& R_\1^4-2 D R_\1^2 R_\2+4 (D-1) R_\1 R_\3+ \left(D^2-3 D+3\right) R_\2^2 -(D-1) D R_\4\,,\cr
{\cal A}^{\rm gf}_\5 &=& e_1 \Big[-2 \left(D^2-2 D+2\right) R_\3 R_\1^2-R_\1 \left(\left(D^2+3 D-3\right) R_\2^2-5 (D-1) D R_\4\right)\cr
&&\qquad+2 D \left(\left(D^2-2 D+2\right) R_\2 R_\3-(D-1) D R_\5\right)+R_\1^5\Big]\cr
&&+ e_2 \Big[\left(D^2-2 D+2\right) R_\2 R_\3-D R_\3 R_\1^2-R_\1 \left(D R_\2^2-3 (D-1) R_\4\right)\cr
&&\qquad-(D-1) D R_\5+R_\2 R_\1^3\Big]\,,\cr
{\cal A}^{\rm gf}_\6 &=&e_1 \Big[3 D^4 R_\3^2-3 D^4 R_\6-6 D^3 R_\3^2+4 D^3 R_\6+D \left(2 D^2-9 D+9\right) R_\2^3-4 D^2 R_\3^2\cr
&&\qquad-2 D^2 R_\6-6 R_\1 \left(D \left(D^2-2 D+2\right) R_\2 R_\3-\left(D^3-2 D^2+2 D-1\right) R_\5\right)\cr
&&\qquad+20 D R_\3^2+15 (D-1)^2 R_\2 R_\4+D R_\6+R_\1^6-10 R_\3^2\Big]\cr
&&+e_2 \Big[-4 R_\1 \left(\left(D^2-D+1\right) R_\2 R_\3-(D-1) D R_\5\right)+\left(D^2-3 D+3\right) R_\2^3\cr
&&\qquad+2 D \left(\left(D^2-2 D+2\right) R_\3^2-(D-1) D R_\6\right)+(D-1) D R_\2 R_\4+R_\2 R_\1^4\Big]\cr
&& + e_3 \Big[D^2 R_\3^2-D^2 R_\6+3 (D-1) R_\5 R_\1+D R_\2^3+D R_\3^2\cr
&&\qquad-3 R_\2 \left(D R_\1 R_\3+(D-1) R_\4\right)+D R_\6+R_\3 R_\1^3-R_\3^2\Big]\cr
&&+e_4 \Big[\left(D^2-2 D+2\right) R_\3^2-2 R_\1 \left(D R_\2 R_\3-(D-1) R_\5\right)\cr
&&\qquad+(D-1) \left(R_\2 R_\4-D R_\6\right)+R_\1^2 R_\2^2\Big]\cr
&&+e_5 \Big[-R_\2 \left(D R_\4+2 R_\1 R_\3\right)+D R_\3^2+R_\2^3+R_\1^2 R_\4\Big]\,.
\eea
Starting from $k=6$, there are quasi-topological Ricci polynomials on special static metric ((\ref{staticmet}) with $h=f$) that also satisfy the $a$-theorem.  For $k=6$, there are two such solutions, given by
\bea
{\cal P}_{\!a\,\6} &=& e_1 \Big[3 D^3 R_\3^2-2 D^3 R_\6+3 \left(D^2-3 D+5\right) R_\2^2 R_\1^2\cr
&&\qquad-6 R_\1 \left(D (3 D-5) R_\2 R_\3-2 \left(D^2-3 D+2\right) R_\5\right)-D \left(D^2-12 D+20\right) R_\2^3\cr
&&\qquad+15 D^2 R_\3^2-30 \left(D^2-3 D+2\right) R_\2 R_\4+6 D^2 R_\6-3 D R_\2 R_\1^4\cr
&&\qquad+4 (3 D-5) R_\3 R_\1^3-60 D R_\3^2-4 D R_\6+R_\1^6+40 R_\3^2\Big]\cr
&&+e_2 \Big[-R_\2 \left(D R_\4+2 R_\1 R_\3\right)+D R_\3^2+R_\2^3+R_\1^2 R_\4\Big]
\eea
When $e_2=-3(D-2)(D+5) e_1$, it becomes quasi-topological on general static metric (\ref{staticmet}), and the resulting Ricci polynomial is
\bea
{\cal U}_{a\,\6} &=& -2 D^3 R_\6+3 R_\1^2 \left(\left(D^2-3 D+5\right) R_\2^2-\left(D^2+3 D-10\right) R_\4\right)\cr
&&-12 R_\1 \left(\left(D^2-4 D+5\right) R_\2 R_\3-\left(D^2-3 D+2\right) R_\5\right)+6 D^2 R_\3^2\cr
&&+6 D^2 R_\6-\left(D^3-9 D^2+29 D-30\right) R_\2^3+3 \left(D^3-7 D^2+20 D-20\right) R_\2 R_\4\cr
&&-3 D R_\2 R_\1^4+4 (3 D-5) R_\3 R_\1^3-30 D R_\3^2-4 D R_\6+R_\1^6+40 R_\3^2\,.
\eea
We have constructed these theories up to the tenth order, as an extension of the works of \cite{quasi4}.  The procedure is straightforward albeit tedious, and we shall not present the details here.  The upshot is that at the sufficiently higher order, which is still quite manageable for Ricci polynomials, one can construct theories with many desired properties.

\section{Including the Riemman tensor}
\label{sec:riemann}

Having studied the $a$-theorem constraint on the Ricci polynomial gravities, we now consider the inclusion of Riemann tensor.  The structures of the Riemann tensor polynomials are much more complicated than those of Ricci polynomials, and we shall consider only the cubic Riemann tensor terms.  There are eight such terms and the Lagrangian is
\bea
{\cal L} &=& \sqrt{-g} \Big(R - 2\Lambda_0 + L_{\rm cubic}\Big)\,,\cr
L_{\rm cubic} &=& e_1 R^3 + e_2 R\,R_{\mu\nu} R^{\mu\nu} + e_3 R^{\mu}_{\nu} R^{\nu}_\rho R^{\rho}_\mu + e_4 R^{\mu\nu} R^{\rho\sigma} R_{\mu\rho\nu\sigma}\cr
&&+ e_5 R R^{\mu\nu\rho\sigma} R_{\mu\nu\rho\sigma} +e_6 R^{\mu\nu} R_{\mu \alpha\beta\gamma} R_{\nu}{}^{\alpha\beta\gamma} +
e_7 R^{\mu\nu}{}_{\rho\sigma} R^{\rho\sigma}{}_{\alpha\beta} R^{\alpha\beta}{}_{\mu\nu}\cr
&&+e_8 R^\mu{}_\nu{}^\alpha{}_\beta R^\nu{}_\rho{}^\beta{}_\gamma R^{\rho}{}_\mu{}^\gamma{}_{\alpha}\,.\label{riemanncubiclag}
\eea
In particular, the Euler integrand of the third order corresponds to taking
\be
\{e_1,e_2,e_3,e_4,e_5,e_6,e_7,e_8\}=\{1, -12, 16, 24, 3, -24, 4, -8\}\,.\label{euler3}
\ee
It is a total derivative in six dimensions and vanishes identically in $D\le 5$.

\subsection{$D=5$}

\subsubsection{Linearized quasi-topological gravity}

We shall first focus on the discussion in five dimensions, where the Euler integrand given by (\ref{euler3}) vanishes identically. The bare and effective cosmological constants are related by
\be
\Lambda_0 = \Lambda_{\rm eff} -\ft{1}{108} \left(400 e_1+80 e_2+16 e_3+16 e_4+40 e_5+8 e_6+4 e_7+3 e_8\right) \Lambda_{\rm eff} ^3\,.
\ee
As in the previous cases, there are additional massive scalar and spin-2 modes in the linearized spectrum of the AdS vacua, and the absence of these modes requires
\bea
20 e_2+12 e_3+7 e_4+80 e_5+20 e_6+24 e_7-3 e_8&=&0\,,\nn\\
320 e_1+76 e_2+20 e_3+17 e_4+48 e_5+12 e_6+8 e_7+3 e_8&=&0\,.\label{nomassive}
\eea
To be specific, the first equation implies the decoupling of the massive spin-2 modes and the resulting theory becomes ghost free.  The second equation above implies the decoupling of the massive scalar mode.  Unlike the previously discussed Ricci polynomials, the inclusion of Riemann tensor implies that causality condition may not be satisfied by the exclusion of both massive modes alone. The linearized theory becomes quasi-topological only after an additional condition is satisfied
\be
40 e_5+8 e_6+12 e_7-3 e_8=0\,.\label{riemannquasicond}
\ee
In other words, when these three conditions are satisfied, we have $\Lambda_{\rm eff}=\Lambda_0$ and the linearized gravity becomes identical to that of Einstein gravity, satisfying the causality constraint of \cite{Camanho:2014apa}, on the vacuum background.

\subsubsection{The $a$ and $c$ charges}

With the inclusion of the Riemann tensor, the $a$ and $c$ charges are no longer equal.  In order to read off both charges, we consider the reduced FG expansion with the metric ansatz where the AdS boundary is $S^2\times S^2$:
\be
ds^2 = \fft{\ell^2 dr^2}{4r^2} + \fft{1}{r} \Big(f_1(r)\, (d\theta_1^2 + \sin^2\theta_1 d\phi_1^2) +
f_2(r)\, (d\theta_2^2 + \sin^2\theta_2 d\phi_2^2)\Big)\,.\label{d5s2s2ans}
\ee
The boundary is at $r=0$, and for small $r$, the functions $f_i(r)$ have the following Taylor expansion
\be
f_i = b_{i0} + b_{i2} r + b_{i4} r^2 + \cdots\,,\qquad \hbox{with}\qquad i=1,2\,.
\ee
The Gauss-Bonnet term gives the topological number of $S^2\times S^2$, whilst the Weyl-squared on the boundary depends on the ratio $\gamma=b_{10}/b_{20}$:
\bea
\sqrt{h} E^\4 =8\,,\qquad \sqrt{h} I^\4 = \fft43\Big(2 + \gamma + \fft{1}{\gamma}\Big)\,.
\label{eires}
\eea

Substituting the ansatz (\ref{d5s2s2ans}) into the Lagrangian, the anomaly term can be read off as the coefficient of the $1/r$ term in the small $r$ expansion.  Varying $(a_2,b_2)$ yields
\be
b_{12}=\ft16(\gamma-2)\ell^2\,,\qquad
b_2 = \ft16(\fft{1}{\gamma}-2)\ell^2\,.
\ee
We find that the anomalous $1/r$ term is given by
\bea
{\cal A} &=&
\fft{1+\gamma^2}{12 \gamma} \Big(\frac{1}{\ell}\big(1200 e_1+240 e_2+48 e_3+48 e_4-40 e_5-8 e_6-36 e_7+21 e_8\big)+\ell^3\Big)\nn\\
&&-\frac{\ell^3}{3}-\frac{1}{3\ell}\Big(1200 e_1+240 e_2+48 e_3+48 e_4+200 e_5+40 e_6+36 e_7+3 e_8\Big)\,.
\eea
It follows from (\ref{d5riccianomaly}) and  (\ref{eires}) that the $a$ and $c$ charges are
\bea
a&=& \ell^3 + \frac{1}{\ell} \left(1200 e_1+240 e_2+48 e_3+48 e_4+120 e_5+24 e_6+12 e_7+9 e_8\right)\,,\nn\\
c &=&\ell^3+ \frac{1}{\ell}\left(1200 e_1+240 e_2+48 e_3+48 e_4-40 e_5-8 e_6-36 e_7+21 e_8\right)\,.
\label{riemannac}
\eea
We see that the condition (\ref{riemannquasicond}) that is necessary for linearized quasi-topological gravity ensures precisely that $a=c$.  In particular, we have $a=c$ automatically when the $e_5, e_6,e_7,e_8$ vanish.  We expect that this is generally true that the condition for linearized quasi-topological gravity, which ensures the causality constraint, is equivalent to the equality $a=c$ in the dual CFT.

\subsubsection{The $a$-theorem}

It follows from (\ref{riemannac}) that the $a$-function takes the form
\be
a(r)=A'^3 + \left(1200 e_1+240 e_2+48 e_3+48 e_4+120 e_5+24 e_6+12 e_7+9 e_8\right) \frac{1}{A'}\,.
\ee
Thus we have
\be
a'=3\Big((400 e_1+80 e_2+16 e_3+16 e_4+40 e_5+8 e_6+4 e_7+3 e_8) A'^4 -1\Big) \fft{A''}{A'^4}\,.
\ee
We follow the same strategy in section \ref{sec:simplenec} and find the NEC is
\bea
0\le -{\cal E}^t_t + {\cal E}^r_r &=& a' A'^4 -3 x \left(A''^3+A'''' A'^2+4 A''' A'^3+16 A'^2 A''^2+4 A' A''A'''\right)\cr
&&-3 y \left(A'''^2+4 A''^3+ A''A''''+4 A' A''A'''\right)\,,
\eea
where the coefficients $(x,y)$ are
\bea
x &=& 320 e_1+76 e_2+20 e_3+17 e_4+48 e_5+12 e_6+8 e_7+3 e_8\,,\nn\\
y &=& 128 e_1+40 e_2+17 e_3+8 e_4+32 e_5+10 e_6+8 e_7\,.
\eea
Thus the validity of the $a$-theorem requires that
\be
x=0=y\,.\label{aconds}
\ee
Note that the condition of $x=0$ is precisely the decoupling of the massive scalar mode.

The combined conditions of linearized quasi-topological gravity (which implies $a=c$) and the $a$-theorem leads to four solutions, given by
\bea
&&e_5=-15 e_1 - 2 e_2 - \ft14 e_4\,,\qquad
e_6 = - 24 e_1 -8 e_2 - \ft92 e_3 - e_4\,,\cr
&&e_7=74e_1 + 13 e_2 + \ft72 e_3 + \ft54 e_4\,,\qquad
e_8=32 e_1 + 4 e_2 + 2 e_3 + e_4\,.
\eea
One of the combination is simply the trivial Euler integrand which vanishes identically in five dimensions, leading to three non-trivial combinations.  Note that the resulting theories are not only linearly quasi-topological on the vacua, but also quasi-topological on the domain wall metrics.

In Ricci polynomial gravities, Einstein metrics, including Schwarzschild black holes ((\ref{staticmet}) with $h=f$) continue to be solutions. The inclusion of Riemann tensor will exclude the Schwarzschild black holes.  In literature, there have been considerable interests in studying the Riemann tensor polynomials that would give rise to Schwarzschild-like black holes ($h=f$). (See, e.g.~\cite{quasi0,quasi1}) For the Lagrangian (\ref{riemanncubiclag}), the results are given by \cite{quasi1}
\bea
&& e_ 1=\ft{1}{648}(-128e_ 6-588e_ 7+9e_ 8))\,,\qquad e_ 2=\ft{1}{54}(92e_ 6+372e_ 7-9e_ 8)\,,
\cr && e_ 3=-\ft{2}{81}(67e_ 6+204e_ 7-18e_ 8)\,,\qquad e_ 4=\ft{1}{27}(-52e_ 6-168e_ 7-9e_ 8)\,,
\cr && e_ 5=\ft{1}{216}(-56e_ 6-156e_ 7+9e_ 8)\,. \label{heqfsol}
\eea
In these combinations the massive modes decouple automatically, and hence the theory is ghost free. In other words, the two conditions (\ref{nomassive}) are satisfied. If one requires further the $a$-theorem conditions (\ref{aconds}), the solutions reduce to
\bea
&& e_ 1=\ft{1}{56}(44e_ 7+15e_ 8)\,,\qquad e_ 2=-\ft{3}{14}(36e_ 7+11e_ 8)\,,\qquad e_ 3=\ft{2}{7}(32e_ 7+9e_ 8)\,.
\cr && e_ 4=\ft{3}{7}(24e_ 7+5e_ 8)\,,\qquad e_ 5=\ft{3}{8}(4e_ 7+e_ 8)\,,\qquad e_ 6=-\ft{3}{7}(20e_ 7+3e_ 8)\,.\label{heqfsol2}
\eea
The trivial Euler integrand corresponds to $2e_7+e_ 8=0$.  The non-trivial combination, which can be obtained by setting $e_7=0$ without loss of generality, was obtained in \cite{quasi1}.  It can easily verified from (\ref{riemannac}) that
\be
\fft{a-c}{c} = \fft{48(2 e_7 + e_8)}{7\ell^4 - 12 (2 e_7 + e_8)}\,.
\ee

    If we instead require $a=c$ (but not the $a$-theorem condition) on the solutions (\ref{heqfsol}),
we have
\bea
&&e_1 = \ft18(4e_7 + e_8)\,,\qquad e_2 = - \ft38 (14e_7 + 3 e_8)\,,\qquad
e_3 = \ft18 (54 e_7 + 11 e_8)\,,\cr
&& e_4 = \ft34 (10e_7 + e_8)\,,\qquad e_5=\ft3{16} (6e_7 + e_8)\,,\qquad
e_6=-\ft3{16} (38 e_7 + 3 e_8)\,.\label{heqfsol3}
\eea
The solutions (\ref{heqfsol2}) and (\ref{heqfsol3}) do not overlap except for the trivial Euler integrand (\ref{euler3}).  Thus the existence of Schwarzschild-like black holes in cubic Riemann theory would violate the causality condition.  Of course, this conclusion was drawn for pure gravity. If it could be embedded in a more fundamental theories, additional degrees of freedom such as higher spin modes may come to rescue the causality, in which case causality condition is relaxed to
\be
\Big|\fft{a-c}{c}\Big|\ll 1\,,
\ee
rather than it being strictly zero.  The higher-order terms should then be treated perturbatively only.

\subsection{The $a$-theorem in general dimensions}

The five-dimensional results obtained in the previous subsection can be generalized to general dimensions. The bare and effective cosmological constants are related by the equation of motion, yielding
\bea
\Lambda_0&=&\Lambda_{\rm eff} + \fft{4(D-6)\Lambda_{\rm eff}^3}{(D-1)^2(D-2)^3}
\Big(D^2 (D-1)^2 e_1+D (D-1)^2 e_2+(D-1)^2 e_3\cr
&&\qquad+(D-1)^2 e_4+2 D (D-1) e_5+2 (D-1) e_6-(2-D) e_8+4 e_7
\Big)\,.
\eea
The decoupling of the massive modes requires that
\bea
&&(D-1) D e_2-(3-3 D) e_3-(3-2 D) e_4+4 (D-1) D e_5+4 D e_6+24 e_7-3 e_8=0\,,\nn\\
&&12 D (D-1)^2 e_1+\left(D^2+8 D-8\right) (D-1) e_2+3 D (D-1) e_3+\left(2 D^2+D-4\right) e_4\nn\\
&&+4 (D+4) (D-1) e_5+4 (2 D-1) e_6+3 (D-2) e_8+24 e_7=0\,.
\eea
The first and second equations correspond to the decoupling of the massive spin-2 and massive scalar modes respectively.  The theory becomes quasi-topological at the linearized level around the AdS vacua provided that an additional constraint is satisfied, namely
\be
2 (D-1) D e_5+2 (D-1) e_6+12 e_7-3 e_8=0\,.
\ee
In this case, we have $\Lambda_0 = \Lambda_{\rm eff}$.  The theory is then consistent with the causality condition on these backgrounds.

Following the same strategy discussed in the earlier sections, and using the curvature results presented in the appendix, we find that the $a$ charge is
\be
a=\ell^{D-2} + 3\Big((D-1)^2 (D^2 e_1 + D e_2 + e_3 + e_4) + 2(D-1)(D e_5 + e_6) +
4 e_7 + (D-2) e_8\Big)\ell^{D-6}\,.
\ee
This leads to an $a$-function
\bea
a(r) &=& 3\Big((D-1)^2 (D^2 e_1 + D e_2 + e_3 + e_4) + 2(D-1)(D e_5 + e_6) +
4 e_7 + (D-2) e_8\Big)A'^{6-D}\cr
&&+ A'^{2-D}\,,
\eea
and its derivative
\bea
a'(r) &=& -\Big[(D-2) + 3(D-6)\Big((D-1)^2 (D^2 e_1 + D e_2 + e_3 + e_4)\cr
&&\qquad\qquad\qquad + 2(D-1)(D e_5 + e_6) + 4 e_7 + (D-2) e_8\Big)\Big]
A'^{1-D} A''\,.
\eea
Adopting the same technique earlier and using the curvature results presented in the appendix, we derive the NEC, given by
\bea
-{\cal E}^t_t + {\cal E}^r_r &=& A'^{D-1} a' -3 y \left(A'''^2+(D-1) A''^3+A'' \left(A''''+(D-1) A'A'''\right)\right)\cr
&&+ x \left(A''^3+(D-1)A'^3 A'''+A'^2 \left(A''''+4 (D-1) A''^2\right)+4A'A''A'''\right),
\eea
where
\bea
x &=& -12 D (D-1)^2 e_1 -(D-1)\left(D^2+8 D-8\right) e_2-3 D (D-1) e_3+\left(-2 D^2-D+4\right) e_4
\cr
&&-4 (D+4) (D-1) e_5-4 (2 D-1) e_6-24 e_7-3 (D-2) e_8\,,\cr
y &=&8 (D-1)^2 e_1+2 D (D-1) e_2 + \left(D^2-2 D+2\right) e_3 +2 (D-1) e_4+8 (D-1) e_5\cr
&&+2 D e_6+8 e_7\,.
\eea
As in the all previous cases, the terms involving a linear $A''$ can be factorized to proportional to $a'$.  The vanishing the coefficients $x$ and $y$ yields the $a$-theorem.  In particular, $x=0$ implies the decoupling of the massive scalar mode.

Combining all the above conditions together, we have
\bea
e_5 &=& -\ft34 D(D-1) e_1 - \ft12 (D-1) e_2 - \ft14 e_4\,,\nn\\
e_6 &=& -\ft32 (D-4)(D-1)^2 e_1 - (D-1)(D-4) e_2 - \ft32 (D-2) e_3 -\ft12 (D-3) e_4\,,\nn\\
e_7 &=& \ft18 \left(3 D^2-6 D-8\right) (D-1)^2 e_1+\ft14 \left(D^2-2 D-2\right) (D-1) e_2\cr
&&+\ft14 \left(D^2-2 D-1\right) e_3+\ft18 (D-3) D e_4\,,\nn\\
e_8 &=& 2 (D-4) (D-1)^2 e_1+(D-4) (D-1) e_2+(D-3) e_3-e_4\,.
\eea
These solutions exclude the Euler integrand (\ref{euler3}) for $D\ge 7$, for which it becomes nontrivial and violates the causality condition, at least for pure gravity theories. When $e_5=e_6=e_7=e_8=0$, we have one combination
\be
\{e_1,e_2,e_3,e_4\}=\{1,(2D-1), 2(D-1), (D-1)(D-2)\}\,,
\ee
corresponding to
\be
K^\3 = R^3 + (2D-1) R R^{\mu}_\nu R^\nu_\mu + 2(D-1) R^\mu_\nu R^\nu_\rho R^\rho_\mu +
(D-1)(D-2) R^{\mu\nu} R^{\rho\sigma} R_{\mu\rho\nu\sigma}\,.
\ee
In addition to satisfying all the constraints considered in this subsection, it also admits Einstein metrics as its vacuum solutions.

\section{Conclusions}

In this paper, we studied properties of Einstein gravity with a cosmological constant, extended by curvature tensor polynomial invariants.  We focused on Ricci polynomials and also discussed the general Riemann cubic polynomials.  In all these theories, the vacua are the maximally-symmetric spacetimes with positive, zero or negative effective cosmological constants. We concentrate our attention on the negative effective cosmological constant.
Since we only considered tensor polynomials rather than their derivatives, the linearized gravities in general contain two massive modes, a scalar and a spin-2, in addition to the usual graviton.

We derive the constraints on the coupling constants of the Ricci polynomials from the considerations for the theories being ghost-free, maintaining causality and exhibiting a holographic $a$-theorem.  We find that all these conditions can be satisfied only at the quartic order and beyond.  The resulting theories are linearized quasi-topological gravities in that the higher-order terms do not contribute to the linearized equation at all on the Einstein metric solutions.  Furthermore, they are quasi-topological on domain wall metrics.  Consequently, the causality conditions are satisfied on these backgrounds. This work is an extension of \cite{quasi4}, and starting at the sixth order, we can construct these theories that are also quasi-topological on both special and general static metrics defined in \cite{quasi4}.

The situation becomes more complicated when Riemann tensor polynomials are included.  Einstein metrics are in general no longer solutions of the theory, and linearized gravity is typically derived on a maximally-symmetric vacuum. We studied the analogous properties on general Riemann cubic terms and derived the analogous constraints.  Excluding the Lovelock combination, there are three solutions that satisfy all the criteria. However, these three combinations do not satisfy the condition that (\ref{staticmet}) with $h=f$ is the only static and spherically-symmetric solutions, if we restrict ourselves to pure cubic gravities.  It is thus of great interest to investigate higher order curvature invariants, where Ricci polynomials are much better understood.

\section*{Acknowledgement}

Y-Z.L.~and H.L.~are supported in part by NSFC grants No.~11475024, No.~11175269 and No.~11235003. J-B.W.~is supported in part by
11575202.  H.L.~and J-B.W.~would like to thank the participants of the advanced workshop ``Dark Energy and Fundamental Theory'' supported by
the Special Fund for Theoretical Physics form the National Natural Science Foundations of China with Grant no.~11447613 for stimulating discussions.
\appendix

\section{Curvature of domain wall metrics}

In this paper, we have extensively made use of the domain wall metrics of the type
\be
ds^2_D = e^{2B(r)} dr^2 + e^{2A(r)} d\Omega_{D-1, \varepsilon}^2\,,\label{dwansatz}
\ee
where $d\Omega_{D-1,\varepsilon}^2$ is the maximally symmetric metric with $\tilde R^{ij}{}_{kl}= \varepsilon (\delta^i_k \delta^j_l - \delta^i_l \delta^j_k)$. The non-vanishing components of the Riemann tensors of the metric (\ref{dwansatz}) are
\be
R^{0i}{}_{0j}=X\,,\qquad R^{ij}{}_{k\ell} = Y (\delta^i_k \delta^j_\ell -
\delta^i_\ell \delta^j{}_k)\,,
\ee
where
\be
X=-e^{-2 B} \left(A''-A' B'+A'^2\right)\,,\qquad
Y=e^{-2 A} \varepsilon -e^{-2 B} A'^2\,.
\ee
Thus for curvature polynomials we considered in this paper, we have
\bea
&&R_{(k)} = (D-1) ((D-2) Y+X)^k+((D-1) X)^k\,,\cr
&&R^{\mu\nu} R^{\rho\sigma} R_{\mu\rho\nu\sigma}=(D-1) ((D-2) Y+X) \left(2 (D-1) X^2+(D-2) Y ((D-2) Y+X)\right)\,,\cr
&&R R^{\mu\nu\rho\sigma} R_{\mu\nu\rho\sigma} = 2 (D-1)^2 ((D-2) Y+2 X) \left((D-2) Y^2+2 X^2\right)\,,\cr
&&R^{\mu\nu} R_{\mu \alpha\beta\gamma} R_{\nu}{}^{\alpha\beta\gamma} =
2 (D-1) \left(D X^3+(D-2) X Y (X+Y)+(D-2)^2 Y^3\right)\,,\cr
&&R^{\mu\nu}{}_{\rho\sigma} R^{\rho\sigma}{}_{\alpha\beta} R^{\alpha\beta}{}_{\mu\nu} =
4 (D-1) \left((D-2) Y^3+2 X^3\right)\,,\cr
&&R^\mu{}_\nu{}^\alpha{}_\beta R^\nu{}_\rho{}^\beta{}_\gamma R^{\rho}{}_\mu{}^\gamma{}_{\alpha} =
(D-2) (D-1) Y \left((D-3) Y^2+3 X^2\right)\,.
\eea

\end{document}